\documentclass[twocolumn]{aastex701}

\usepackage{graphicx}%

\begin{document}

\title{Discovery of minute-scale polarization angle rotation in a supermassive black hole jet}

\author[0009-0005-7962-6296]{Aristeidis Polychronakis}
\altaffiliation{}
\affiliation{Department of Physics, University of Crete, GR-70013 Heraklion, Greece}
\affiliation{Institute of Astrophysics, Foundation for Research and Technology-Hellas, GR-70013, Greece}
\affiliation{Department of Astronomy $\&$ Astrophysics, 525 Davey Laboratory, The Pennsylvania State University, University Park, PA 16802, USA}
\email[show]{axp6351@psu.edu}  

\author[0000-0001-9200-4006]{Ioannis Liodakis}
\affiliation{Institute of Astrophysics, Foundation for Research and Technology-Hellas, GR-70013, Greece}
\affiliation{Institute of Astrobiology, University of Crete, GR-70013, Heraklion, Greece}
\email{liodakis@ia.forth.gr}  

\author[0000-0003-0611-5784]{Dmitry Blinov}
\affiliation{Department of Physics, University of Crete, GR-70013 Heraklion, Greece}
\affiliation{Institute of Astrophysics, Foundation for Research and Technology-Hellas, GR-70013, Greece}
\email{dmitriy.blinov@gmail.com}  

\author[0009-0004-4933-0956
]{Anastasia Glykopoulou}
\affiliation{Department of Physics, University of Crete, GR-70013 Heraklion, Greece}
\affiliation{Institute of Astrophysics, Foundation for Research and Technology-Hellas, GR-70013, Greece}
\email{ph5507@edu.physics.uoc.gr}  

\author[0000-0001-6314-9177]{Sebastian Kiehlmann}
\affiliation{Institute of Astrophysics, Foundation for Research and Technology-Hellas, GR-70013, Greece}
\email{skiehlmann@mail.de}  

\author[0000-0002-6514-7033]{Karan Pal}
\affiliation{Department of Space, Earth and Environment, Chalmers University of Technology, Gothenburg, Sweden}
\affiliation{Dr. Karl Remeis-Observatory and Erlangen Centre for Astroparticle Physics, Friedrich-Alexander Universität Erlangen-Nürnberg, Sternwartstr. 7, 96049 Bamberg, Germany}
\email{karan.pal.3107@gmail.com}  

\author[0000-0001-6757-3098]{Georgios F. Paraschos}
\affiliation{Finnish Centre for Astronomy with ESO, University of Turku, 20014 Turku, Finland}
\affiliation{Aalto University Mets\"{a}hovi Radio Observatory, Mets\"{a}hovintie 114, FI-02540 Kylm\"{a}l\"{a}, Finland}
\email{gfpara@utu.fi}  

\author[0009-0003-8342-4561]{Lena Debbrecht}
\affiliation{Max-Planck-Institut f\"{u}r Radioastronomie, Auf dem H\"{u}gel 69, Bonn, D-53121, Germany}
\email{ldebbrecht@mpifr-bonn.mpg.de}  

\author[0000-0002-2381-4184]{Swati Ravi}
\affiliation{MIT Kavli Institute for Astrophysics and Space Research, Cambridge, MA 02139, USA}
\email{swatir@mit.edu}  

\author[0000-0002-6492-1293]{Herman L. Marshall}
\affiliation{MIT Kavli Institute for Astrophysics and Space Research, Cambridge, MA 02139, USA}
\email{hermanm@space.mit.edu}

\begin{abstract}
Relativistic jets from supermassive black holes oriented towards our line of sight called blazars have puzzled the astronomical community for decades. Their fast jets and preferential alignment creates a fog of relativistic effects that obscures their true properties.  Optical polarimetry - tracing the uniformity, direction, and evolution of the magnetic field in the jets, has revealed a unique phenomenon to blazars; coherent changes of the magnetic field  manifesting as rotations of the polarization angle ($\rm\Psi$). The origin of those rotations has been debated for decades with several proposed models to explain them. We detected an extremely fast - minute time-scale -  rotation in the polarization angle of $\sim$136$^\circ$ over the span of about 80 minutes, making it the fastest known rotation by almost an order of magnitude. Our optical polarization observations combined with radio very long baseline interferometry allow us to reject different models of particle acceleration and strongly point to relativistic effects and the interaction of standing and moving shock waves as the origin of the extreme polarization variations in jets.
\end{abstract}



\section{Introduction} 

Blazars show bright and variable multimessenger emission, showing unique phenomenology among other non-blazar active galactic nuclei \citep{Blandford2019,Hovatta2019}. In the optical, the synchrotron emission from the jet is highly polarized reaching a polarization degree ($\rm\Pi$) of up to 50\% \citep{Agudo2025}, fairly close to the theoretical maximum of about 70\% for a uniform magnetic field. It typically shows erratic variability in both the optical polarization degree and polarization angle on time-scales as short as minutes \citep{Liodakis2024, Poly_2025}. However, the magnetic field in the jets often undergoes coherent changes of its line-of-sight average direction, as traced by the polarization angle ($\rm\Psi$), accompanied by high-energy emission  \citep{Blinov2018}, indicating the simultaneous energization of particles in the plasma of the jet. The origin of these $\rm\Psi$ rotations is still debated, with several different proposed models to explain them. Those include shock waves moving in the jet \citep{Marscher1985,Marscher2008,Marscher2010,Liodakis2020, Paraschos2025a, Paraschos2025b}, kink instabilities \citep{Dong2020,Jorstad2022}, magnetic reconnection \citep{Zhang2018,Hosking2020,Zhang2020}, and random walks of the magnetic field due to turbulence \citep{Marscher2014,Kiehlmann2016}.   Phenomenologically, the RoboPol survey \citep{Blinov2021} as well as other monitoring programs and single sources studies (e.g., \citealp{Larionov2008,Covino2015,Jermak2016,Raiteri2017-II,Marscher2017,Raiteri2021-II}) have observed a wide variety of behaviors in different sources. Apart from the connection to $\gamma$-ray activity \citep{Blinov2018} and an on average drop of the polarization degree during rotations \citep{Blinov2016,Blinov2016-II} no other pattern has emerged. Sources show both clock-wise and counterclock-wise rotations (even by the same source) with typical timescales of the order of days and amplitudes from 90$^\circ$ all the way to 720$^\circ$ \citep{Marscher2010}. Recent observations from the Imaging X-ray Polarimetry Explorer \citep{Weisskopf2022,Soffitta2023} have also revealed $\rm\Psi$ rotations \citep{DiGesu2023,Maksym2025,Pacciani2025}, making rotations a multiwavelength phenomenon. 

One of the difficulties in differentiating between models for $\rm\Psi$ rotations is the inherent 180$^\circ$ ambiguity of $\rm\Psi$. Studying a large sample of blazars with the RoboPol polarimeter at the Skinakas observatory in Crete, Greece, we found that even a few day cadence is not enough to correctly resolve the variations of the magnetic field \citep{Kiehlmann2021}. This would suggest that the magnetic field can likely produce $\rm\Psi$ rotations within a single night, which could provide strong constrains on the mechanism responsible. For this reason we started a new monitoring program at the Skinakas observatory in Crete called the Black hOle Optical polarization TimE-domain Survey (BOOTES). Here we present the discovery of the first intra-night rotation and discuss its likely origins. In section \ref{sec:data} we describe the observations, in section \ref{sec:origin} the possible interpretation for the nature of the extremely fast rotation, and conclude in section \ref{sec:concl}. In section \ref{sce:data_availability} we present the data availability of this work.

\section{Observations and data analysis}\label{sec:data}

\begin{figure*}
  \centering
  \includegraphics[width=\linewidth]{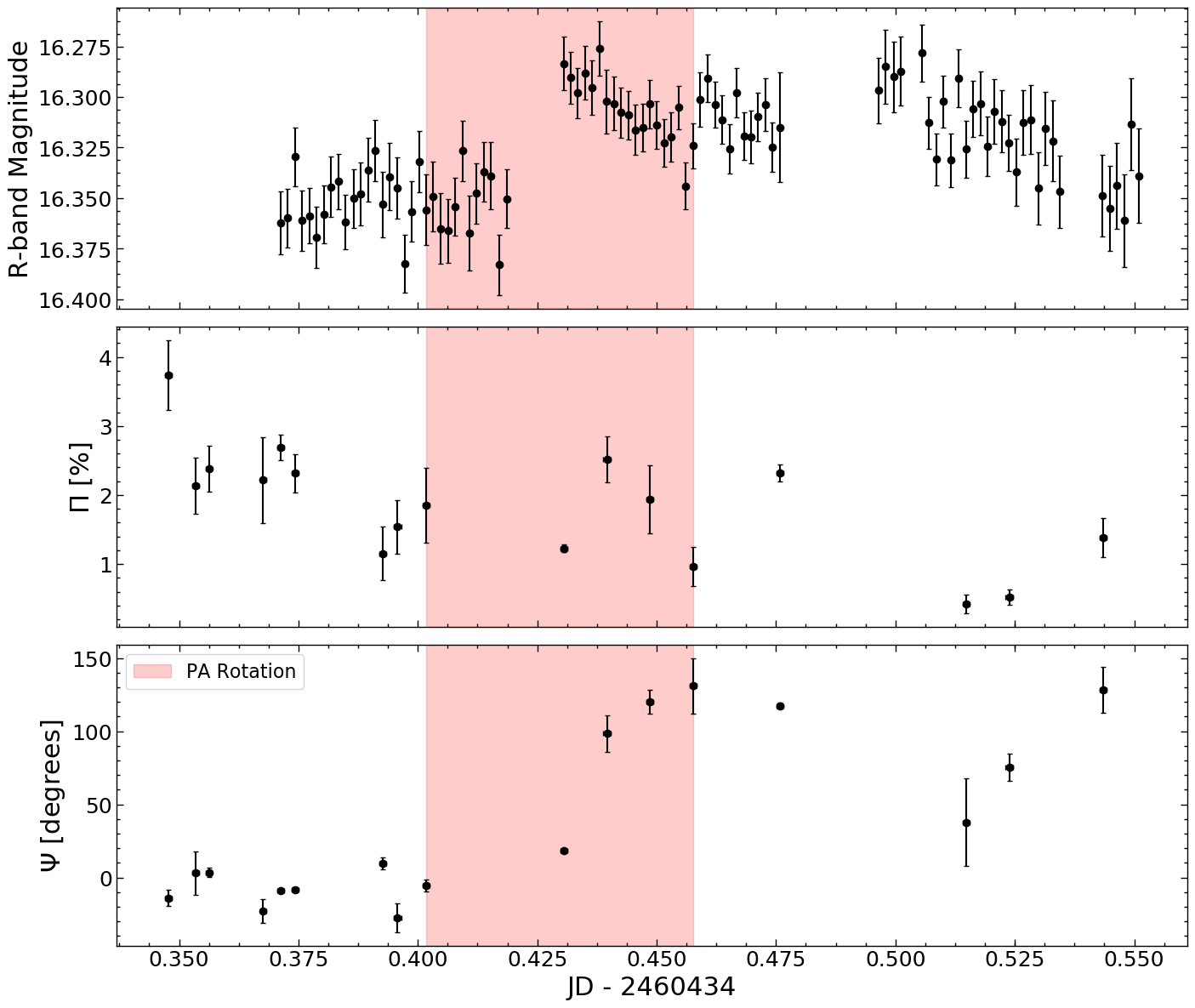}
  \caption{Optical intra-night observations of the blazar PKS~1510-089. The top panel presents the R-band flux density (mJy), the middle panel shows the polarization degree (\%), and the bottom panel displays the polarization angle (degrees). The red-shaded regions indicate the time interval during which the polarization angle rotation takes place. }
  \label{plt:mag_PD_PA}
\end{figure*}

\subsection{Optical Observations}
The optical polarization data used in this work have been presented as part of the sample study in \cite{Poly_2025} and are publicly available via the Harvard
Dataverse (see section \ref{sce:data_availability}). The observations took place on the night of 3 to 4 May 2024, using the RoboPol polarimeter at the Skinakas Observatory in Crete, Greece \cite{Pavlidou2014,Blinov2021}. The data were reduced with the automatic pipeline described in \citep{King2014,Panopoulou2015,Blinov2021}. Adjacent observations were binned in pairs to improve the signal to noise. The observations are shown in Fig. \ref{plt:mag_PD_PA}, where the top panel presents the magnitude of the R-band, the middle panel shows the polarization degree ($\Pi$), and the bottom panel displays the polarization angle ($\Psi$).  We observe a polarization angle variation of approximately $136^\circ$ over a time span of 0.056 days, which is about 80 minutes shown in Fig. \ref{plt:mag_PD_PA} with the red-shaded regions. These values are obtained by directly measuring, in Fig. \ref{plt:mag_PD_PA} the red-shaded area, the time separation between the two boundary points and the corresponding change in $\Psi$. This would suggest a rotation rate of approximately 2450$^\circ$/day almost an order of magnitude faster than the previously fastest known rotation of $\sim360^\circ$/day \cite{MAGIC2018-II}. We find a fairly low degree of polarization  between 1\% and 4\% that fluctuates about 2\%, and remains stable within uncertainties throughout the $\rm\Psi$ rotation. At the same time, there is an increase in brightness at the time of the rotation, suggesting a low-amplitude flare consistent with the expectations from shocks. We apply the $\rm\Psi$ rotation detection algorithm described in \cite{Glykopoulou2026} to confirm the presence of the rotation. The monitoring data were filtered to retain only statistically reliable measurements (\(\mathrm{\Pi}/\mathrm{\sigma}_\mathrm{\Pi} \geq 3\)), and the $\rm\Psi$ values were unwrapped through an error--weighted adjustment procedure to resolve the intrinsic \(180^{\circ}\) ambiguity. Bayesian Blocks segmentation were applied to the adjusted $\rm\Psi$ curve, yielding statistically homogeneous intervals from which local extrema were extracted. The rotation satisfied all selection criteria, including the requirement of sufficient internal structure and data density, and was validated by both a one-sample \(t\)-test (\(\rm p_{t-test} = 3.10 \times 10^{-8}\)) and a binomial test (\(\rm p_{\text{binomial}} = 1.53 \times 10^{-5}\)). The results obtained from this more sophisticated analysis—both the duration of the detected rotation and the total change in $\Psi$—are consistent with those derived from our simpler approach.

\subsection{Very Long Baseline Array data analysis}

\begin{figure*}
  \centering
  \includegraphics[width=\linewidth]{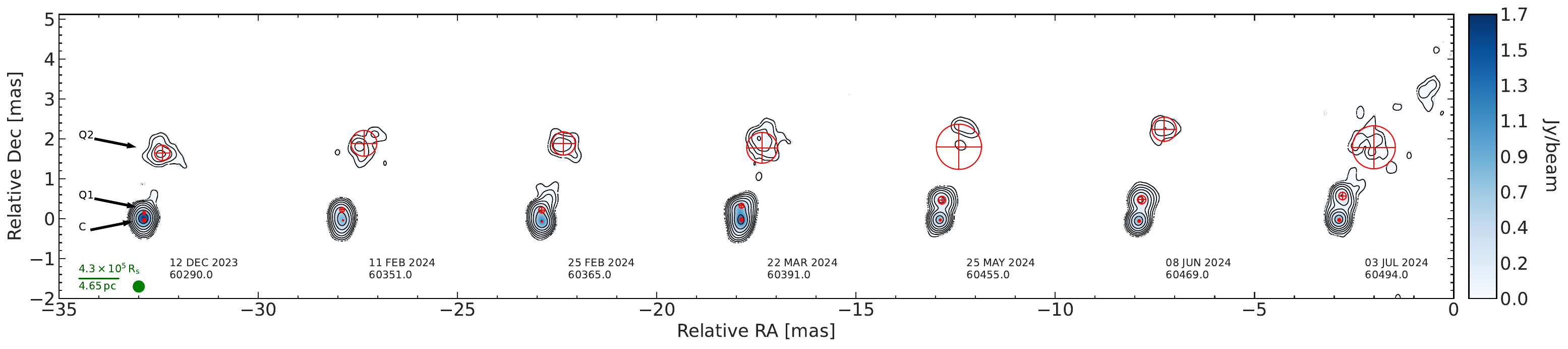}
  \caption{Stokes I images of the VLBA observations of 1510-089. The figure shows the epochs close in time to the observed polarization angle rotation in the optical (MJD~60433). The common colour bar indicates the flux density in Jansky per beam for each epoch and the contours are displayed as 0.25, 0.5, 1, 2, 4, 8, 16, 32, and 64\% of the total VLBI flux density peak of $I_\textrm{max}=1.74\,\textrm{Jy/beam}$. The cut-off is set at the level of the RMS, which is of the order of $8\,\textrm{mJy/beam}$. The observing date is indicated bottom right of each observation, the common circular convolving beam in indicated with the dark green disc (0.28\,mas), and the green bar corresponds to a projected distance of $4.3\times10^{5}\,R_\textrm{S}$. The red circles and crosses indicate the Gaussian model-fit components. The compact region is best modeled by two components (C and Q1) with the distance between them clearly increasing with the passage of time. Q1 appears to break free between March and May 2024.}
  \label{plt:vlbi}
\end{figure*}

\begin{figure}
  \centering
  \includegraphics[width=0.8\linewidth]{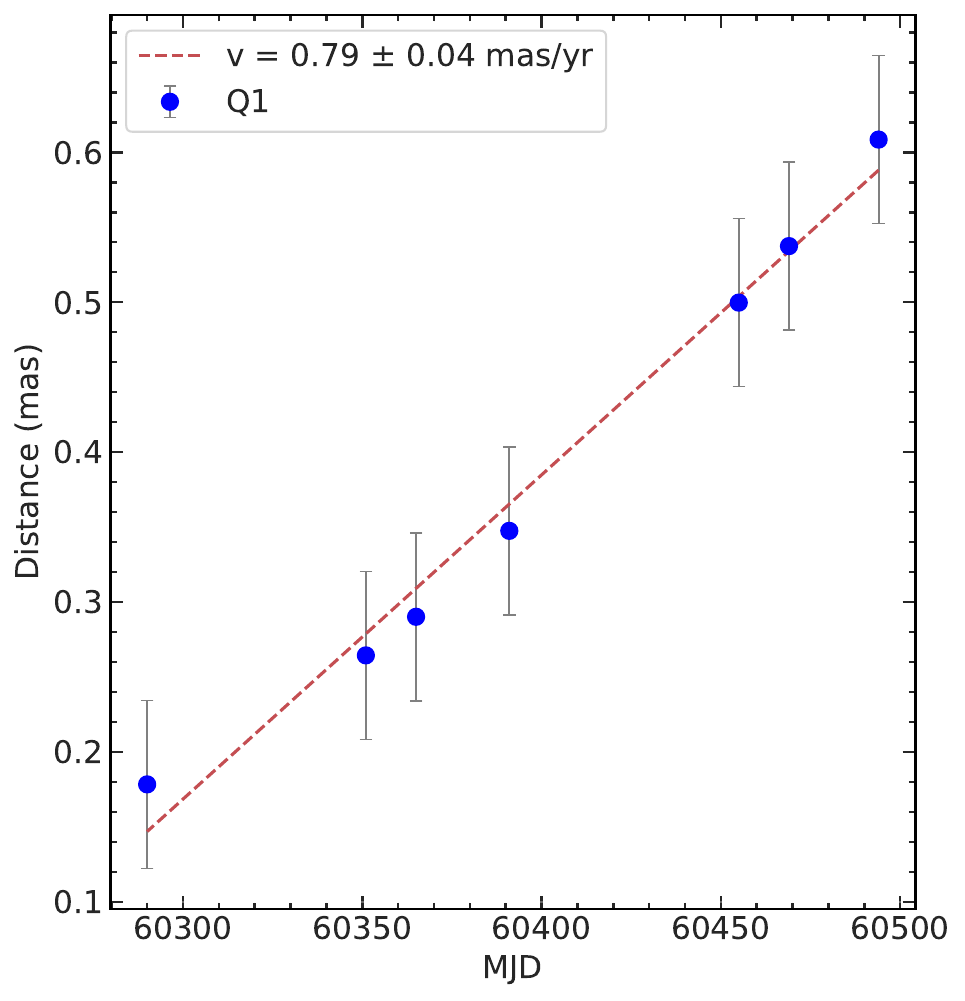}
  \caption{Distance from the core in mas versus modified Julian date for the jet feature Q1. A linear motion is assumed and fitted to the data (dashed red line), yielding a velocity of $v=0.79\pm0.04\,\textrm{mas/year}$ and a date of ejection of $\textrm{MJD} = 60222.10 \pm 10.65$. The positional uncertainties are of the order of 1/5th of the beam size \citep[see also][]{Paraschos22}.}
  \label{plt:VLBI_velocity}
\end{figure}

We use Very Long Baseline Array observations at 43~GHz from the BEAM-ME program. Details about the observations, data calibration, and imaging can be found in \cite{Jorstad2017}. During the time frame of interest, 1510-089 was observed seven times (the corresponding modified Julian dates range from 60290 to 60494), as shown in Fig.~\ref{plt:vlbi}.  The epochs are aligned at the core region, which is the central compact region that we assume to be stationary, based on historical data of the source \citep[e.g.,][]{Jorstad05, Marscher10, Orienti11}. We used the aforementioned publicly available total intensity images and conducted a geometrical modeling of their structure. Specifically, we used two-dimensional, circular Gaussian functions to approximate the source structure through the months, in a procedure known as model-fitting. This approach allows for a simplified representation of the complex jet structure, enabling us to effectively track the emission and trajectory of new jet features \citep[see also][for more details]{Paraschos24a, Paraschos24c}. The best fit model required three model-fit components, labeled `C' (for core), `Q1', and `Q2' (see Fig.~\ref{plt:vlbi}). It is evident that Q1 is moving away from the core, finally splitting away from it contemporaneously to our reported PD rotation (between March and May 2024). Extrapolating to the origin (Fig. \ref{plt:VLBI_velocity}) we find that the component was traveling through the core structure since about October 10 (MJD = 60222.10$\pm$10.65). We estimate the velocity of the component to be 0.790 $\pm$ 0.044 mas/yr, which for an angular diameter distance of 1049.22 Mpc, translates to an apparent velocity of  $\beta_{app}=17.840\pm0.987$c. Assuming a viewing angle at 43~GHz of $\theta_{43}=1.4^\circ$ from \cite{Weaver2022} we estimate a Doppler factor of $\delta_{43}=42.19\pm1.38$. This value exceeds both the variability Doppler factor reported at 15~GHz ($\delta_{var}=32\pm8$, \citealp{Liodakis2018-II}) as well as previous VLBA measurements at 43~GHz of $\delta_{VLBA}=37\pm2.7$ \citep{Weaver2022}.

\subsection{$\gamma$-ray observations}
To explore the potential $\gamma$-ray emission of the source during the polarization angle rotation we use the publicly available observations from the Large Area Telescope (LAT) on board the {\it Fermi} gamma-ray space telescope light curve repository \citep{repository2023}\footnote{\url{https://fermi.gsfc.nasa.gov/ssc/data/access/lat/LightCurveRepository/}}. Since late 2021 the source has entered a quiescent state with no visible flaring activity within the sensitivity of {\it Fermi}-LAT ($\rm 10^{-8} - 10^{-7}~ph/cm^2/s$ in the 0.1-100 GeV band), while historically the brightest state was $2\times$ orders of magnitude higher ($\rm \sim10^{-5}~ph/cm^2/s$). In May of 2024, the flux was approximately $\rm\sim10^{-7}~ph/cm^2/s$ with no evidence of flaring activity even at low level red on timescales of a few days centered around the event.

\section{Origin of the intranight rotation} \label{sec:origin}

\begin{figure}
  \centering
  \includegraphics[width=\linewidth]{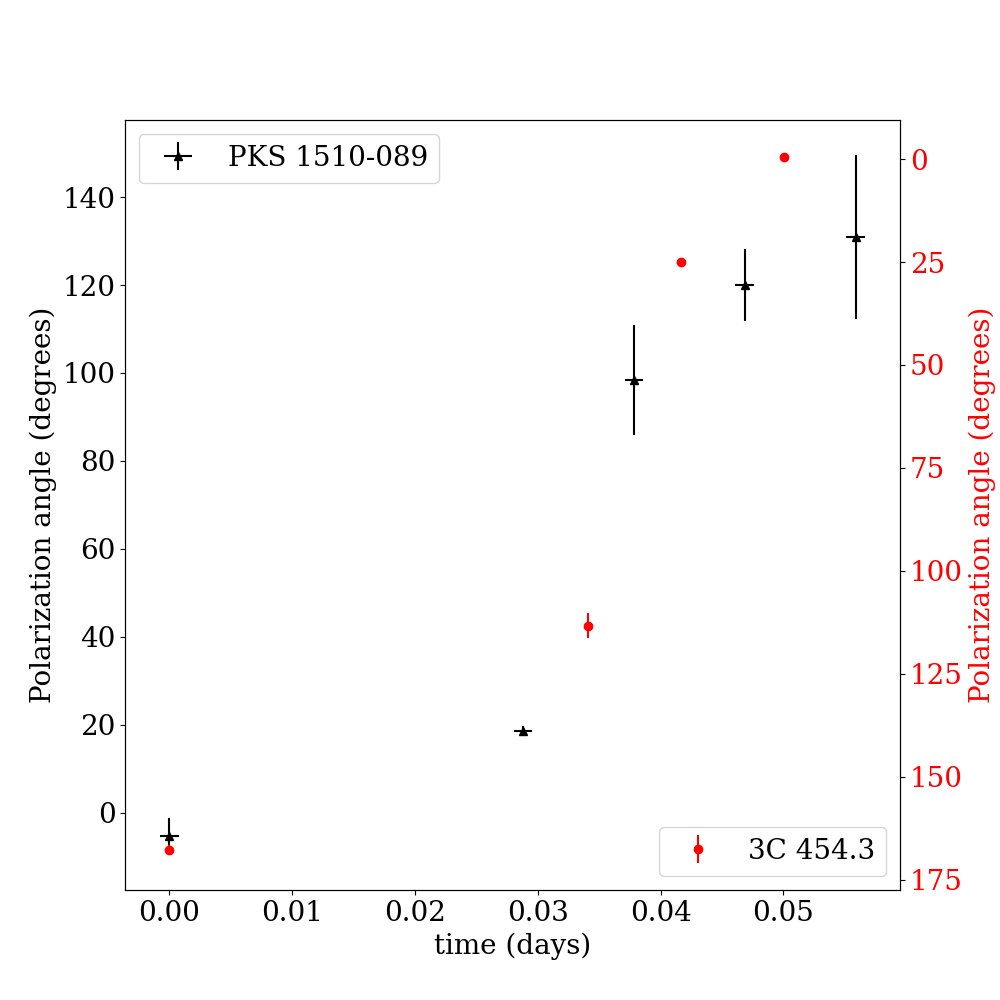}
  \caption{Optical intra-night $\Psi$ of the blazar PKS~1510-089 (black) compared to the relativistic-effect- and redshift- corrected $\Psi$ light curve of 3C~454.3 (red) from \cite{Liodakis2020}. The axis for 3C~454.3 has been inverted so that both rotations follow the same direction. The duration of the 3C~454.3 rotation has been compressed by a factor of 0.12.}
  \label{plt:PA_comparison}
\end{figure}

Using observations from {\it Fermi}-LAT, we do not find evidence for a long-term $\gamma$-ray flaring (within a few days) Instead, we only find $\gamma$-ray upper limits and overall the source was in a low $\gamma$-ray state compared to the long-term light curve as seen by {\it Fermi}-LAT. However, the short duration of the event adds complexity to evaluating the $\gamma$-ray state of the source. First of all, the sensitivity of {\it Fermi}-LAT is not sufficient to detect variations on such short timescales. Second of all, at the time of the rotation, the source was located 65 - 70 degrees from the center of the field of view,  which is the outer limit of useful LAT response as per the {\it Fermi} documentation. Therefore, even if there was a $\gamma$-ray brightening during the rotation, it would have been challenging to detect it. Instead, there is a clear, although low amplitude, brightening of the source in our optical observations during the rotation.  Optical and $\gamma$-ray emission has been shown to correlate using large samples of blazars \citep{Liodakis2018,Liodakis2019,deJaeger2023} suggesting co-spatiality of the emission. 
Therefore, it is not unlikely that there was a low-amplitude $\gamma$-ray brightening during the rotation as is the typical behavior \citep{Blinov2018}, but given the visibility and sensitivity limitations, we are not able to detect.

In addition to the optical and $\gamma$-ray observations, we use publicly available radio Very Long Baseline Array (VLBA) data from the BEAM-ME program \citep{Weaver2022}. This was motivated by previous works that indicate a potential connection between $\rm\Psi$ rotations and the large-scale geometry or physical changes in the jet, for example, jet bending \cite[e.g.,][]{Abdo2010}, or the ejection of new components  \cite[e.g.,][]{Marscher2008}. We detect the presence of a new jet component (Fig. \ref{plt:vlbi}) that clearly separated from the core region sometime between March 22, 2024 and May 25, 2024. The component was traveling through the core region since October 2023 with a high apparent velocity leading to the highest Doppler factor reported so far ($\delta=42\pm1$) for the source. The core region in blazar jets has been suggested to be a system of unresolved standing shocks \cite[e.g.,][and references therein]{Jorstad2010}. Therefore, it is highly likely that the optical rotation is connected to the interaction of new jet component with the standing shock complex during its exodus from the core region. 

Magnetic reconnection has been proposed to produce large swings of the polarization angle that can take place on short time scales similar to what we observe  \citep{Zhang2018,Hosking2020,Zhang2020}. However, the $\rm\Psi$ rotation through magnetic reconnection are usually accompanied by bright flares across the electromagnetic spectrum contradicting our observations. Magnetic reconnection events are also not expected to be accompanied with ejections of new jet components. The passage of moving shock waves through standing shocks can trigger kink instabilities that will produce fast periodic variations of the polarization degree and angle \citep{Jorstad2022}. We tested our observations for periodic variability using the generalized Lomb-Scargle periodogram (see Appendix \ref{sec:periodicity}) and do not find any statistically significant periodicity. Finally, random walks of the polarization angle likely due to turbulence can produce short time-scale $\rm \Psi$ swings, however, the nature of the mechanism will likely produce much higher variability in $\Pi$ which is not seen in our observations. Again, random processes are unlikely to be related to ejections of new features.

To better understand the nature of the $\rm\Psi$ rotation, we use observations from a different blazar called 3C~454.3 that has also shown a similar, but much slower rotation in 2014 \citep{Liodakis2020}. The rotation was accompanied by flaring in optical, $\gamma$-rays, brightening in X-rays and the ejection of a new jet component, similar to this rotation.  Prior to the rotation we were able to track the new jet component (K14) several months before emerging from the radio core by observing an apparent inward motion of the standing feature nearest to the radio core. The observed $\rm\Pi$ and $\rm\Psi$ variability, as well as the optical spectral variations strongly pointed to a shock moving in the jet. The direct comparison of the PKS~1510-089 and 3C~454.3 clearly shows the former to be much faster (1.3 hours versus 12 hours). However, PKS~1510-089 has a much higher Doppler factor and lower redshift than 3C~454.3 \citep{Liodakis2018-II}. Correcting for the relativistic effects (see Appendix \ref{sec:relativity}) we find a remarkable similarity between the amplitude, shape, and time-scales of the two events (Fig. \ref{plt:PA_comparison}) clearly pointing to a common origin. In the case of 3C~454.3, there was a large amplitude variation of the polarization degree, including a drop to near-zero during the rotation. We do not observe such large variation, however, it is possible that our observations were not sensitive enough to capture the full variation of the polarization degree on such short timescales, resulting in a more sparse light curve. Therefore, we cannot exclude the possibility of a similar behavior in the rotation of PKS~1510-089. Since there is little doubt on the nature of the rotation in 3C~454.3, it would suggest that even the fastest variations of the polarization in blazars can be attributed to a combination of relativistic effects and the superposition of interacting polarized components, likely standing and moving shocks \citep{Cohen2020}.

\section{Conclusions}\label{sec:concl}
We have discovered the fastest coherent variation of the magnetic field in a blazar jet, which manifested as a rotation of the optical $\rm\Psi$. Contemporaneous radio observations revealed the emergence of a new jet component, which combined with archival polarization observations, strongly point to shocks propagating in the jets as the origin of even the fastest known $\rm\Psi$ rotations. Shocks have been identified in previous works as the driving mechanism of $\rm\Psi$ rotations \citep{Marscher2008,Marscher2010}, however they are typically associated with much slower events lasting days. We have demonstrated that shocks propagating in the jets are responsible for even the fastest $\rm\Psi$ rotations due to relativistic effects significantly shortening the observed timescales. Results from the BEAM-ME \citep{Weaver2022} and MOJAVE \citep{Lister2021} programs suggest that the ejections of new components are not uncommon. Future intranight polarization observations of large samples of blazars combined with VLBI experiments can further establish the connection between extreme rotations and shocks moving in the jet.

\section{Data Availability}\label{sce:data_availability}

The intra-night optical polarization measurements used in this work
were obtained by \citet{Poly_2025} and are publicly available
via the Harvard Dataverse
\citep{Polychronakis2025_data}. We have also created a new Harvard Dataverse database containing
the full set of polarization data used in this work for PKS 1510-089: Stokes parameters from
\citet{Poly_2025}, together with the simultaneous R-band photometric
magnitudes presented here, available at \cite{Poly_data_2026}.

\begin{acknowledgments}
The authors thank the anonymous referee for insightful comments that helped improve this work. A.G., S.K., and I.L. were funded by the European Union ERC-2022-STG - BOOTES - 101076343. D.B. acknowledges support from the European Research Council (ERC) under the Horizon ERC Grants 2021 programme under grant agreement No. 101040021. G.P. is supported by the European Research Council advanced grant “M2FINDERS - Mapping Magnetic Fields with INterferometry Down to Event hoRizon Scales” (Grant No. 101018682). Views and opinions expressed are however those of the author(s) only and do not necessarily reflect those of the European Union or the European Research Council Executive Agency. Neither the European Union nor the granting authority can be held responsible for them. This study makes use of VLBA data from the VLBA-BU Blazar Monitoring Program (BEAM-ME and VLBA-BU-BLAZAR; \url{http://www.bu.edu/blazars/BEAM-ME.html}), funded by NASA through the Fermi Guest Investigator Program. The VLBA is an instrument of the National Radio Astronomy Observatory. The National Radio Astronomy Observatory is a facility of the National Science Foundation operated by Associated Universities, Inc.
\end{acknowledgments}

\facilities{Skinakas Observatory, Very Long Baseline Array}



\appendix



\subsection*{Periodicity analysis}\label{sec:periodicity}

\begin{figure}
  \centering
  \includegraphics[width=\linewidth]{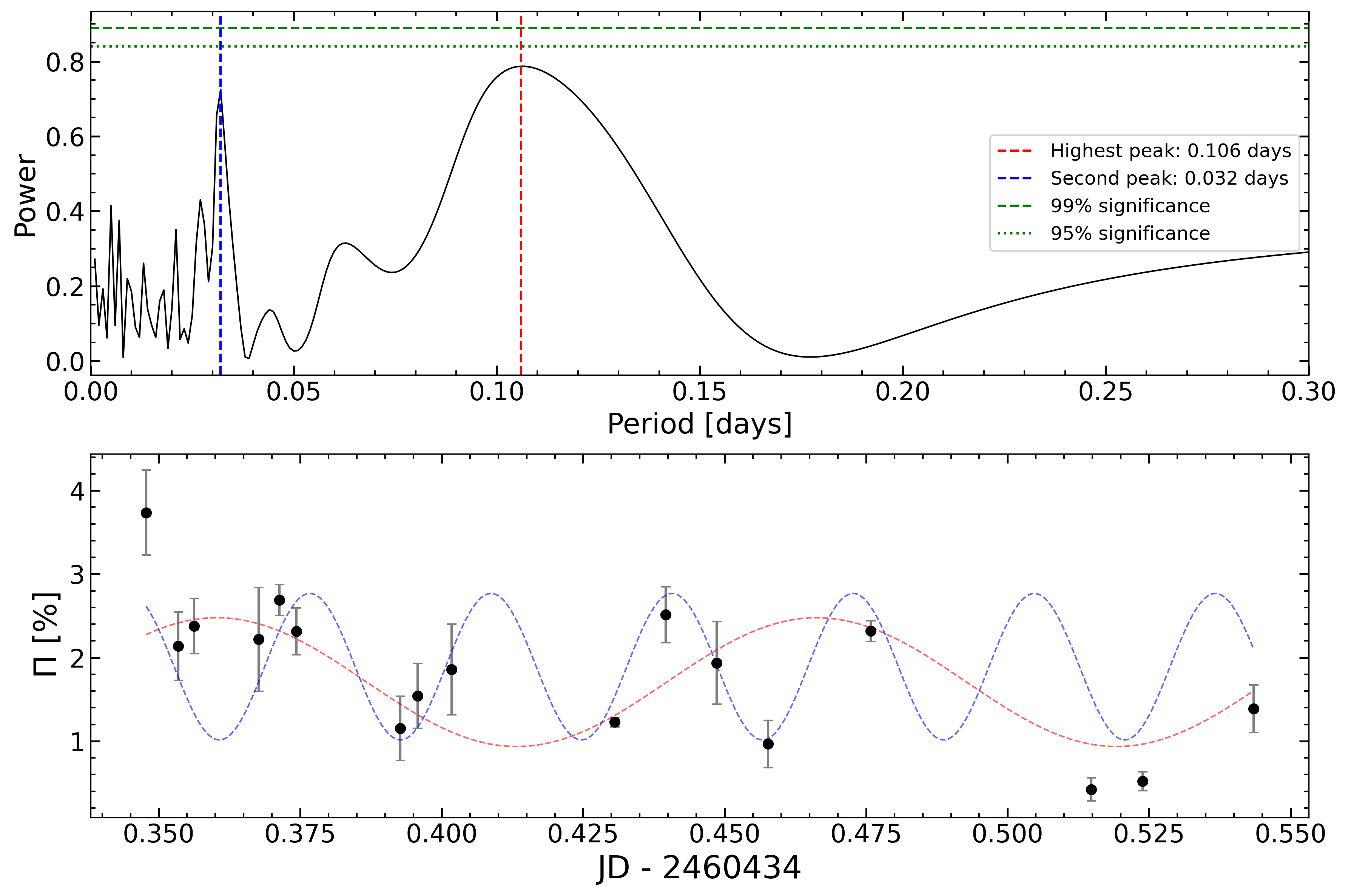}
  \caption{Periodicity analysis of the polarization degree ($\Pi$). The top panel shows the generalized Lomb–Scargle periodogram. The two vertical dashed lines mark the two most prominent periods, while the horizontal green lines indicate the significance thresholds for the detected signals. The bottom panel displays the polarization degree as a function of time. The red and blue dashed curves correspond to sine fits with the periods identified in the top panel (red: 0.106 days; blue: 0.032 days).}
  \label{periodicity}
\end{figure}

In a kink instability scenario, we expect to detect periodic fluctuations in the brightness, $\rm\Pi$, and $\rm\Psi$ related to the growth time-scale of the kink \citep{Dong2020}. We use the generalized Lomb-Scarge periodogram and use simulated light curves generated assuming white-noise to statistically evaluate the significance of a periodic signal (Fig. \ref{periodicity}).

We find two dominant frequencies in the $\rm\Pi$ light curve at 0.106 days (or 152.6 minutes) and 0.032 days (or 46.1 minutes). However, none of the frequencies are statistically significant, suggesting that any periodic pattern is likely a random chance  coincidence  due to random noise. This lack of statistical significance indicates that the detected peaks in the periodogram do not exceed the thresholds corresponding to the 95\% or 99\% confidence levels. In other words, the probability that such peaks could arise purely from stochastic fluctuations in the data is higher than the accepted false-alarm thresholds, and therefore no firm claim of periodicity can be made. While these frequencies may hint at possible underlying variability timescales, we cannot claim that a true periodic signal is present.

\subsection*{Relativistic effect correction}\label{sec:relativity}

In order to correct the relativistic effects and bring the two rotations in the same frame we need to account for the difference in redshift and Doppler factor between PKS~1510-089 and 3C~454.3. The observed time-scale is related to the intrinsic as,
\begin{equation}
 \Delta{t}_{obs} = \frac{1+z}{\delta} \Delta{t_{int}},
\end{equation}
where $\Delta{t}_{obs}$ is the observed time, $\Delta{t_{int}}$ is the intrinsic time, $z$ is the redshift, and $\delta$ is the Doppler factor. We can correct the relativistic and redshift effect difference by,
\begin{equation}
\Delta{t_{int,1}} = \Delta{t_{int,2}}  \Rightarrow \frac{\delta_1}{1+z_1} \Delta{t_{obs,1}} = \frac{\delta_2}{1+z_2} \Delta{t_{obs,2}} \Rightarrow\Delta{t_{obs,1}} =\frac{(1+z_1)\delta_2}{(1+z_2)\delta_1} \Delta{t_{obs,2}},
\end{equation}
where the subscripts (1) and (2) denote PKS~1510-089 and 3C~454.3 respectively. PKS~1510-089 is at a redshift of $z_1=0.36$ while 3C~454.3 is at $z_2=0.859$. We use the Doppler factors from \cite{Liodakis2018-II}, $\delta_1=32.14_{-7.97}^{+8.07}$ and $\delta_2=26.61_{-2.97}^{+6.28}$, to derive a distribution of correction factors through 100,000 random draws. We estimate a median correction factor of 0.60 with a standard deviation of 0.24. This would suggest that the timescales of the rotations are consistent within 2$\sigma$. Using the Doppler factor of the component estimated through the movement of the newly detected component (see VLBA section above) we find a correction factor within uncertainties (0.45$\pm$0.08). Even using the values for the apparent velocity of the $C_{in}$ component from \cite{Liodakis2020} to re-calculate the Doppler factor for 3C 454.3 we come to similar results. Such large variations in the relativistic effects have been observed before in blazars \cite[e.g.,][]{Raiteri2017,Raiteri2017-II,Kovalev2026} . Specifically for 3C~454.3, two consecutive components (K9, K10) found in \cite{Jorstad2013} showed Doppler factors of 27 and 54 respectively. The component in \cite{Liodakis2020} discuss here (K14) showed a Doppler factor of 22 showcasing how beaming can change within an individual blazar from one outburst to another. Therefore, given the uncertainties on the relativistic effects, using three different independent estimates we can fully account the timescale differences between the two events.

\bibliography{sample701}{}
\bibliographystyle{aasjournalv7}



\end{document}